\documentclass[12pt]{iopart}

\pdfoutput=1

\usepackage{iopams} 
\usepackage{graphicx}
\usepackage{amssymb}  
\usepackage{color}
\usepackage{cite}

\usepackage{setstack}
\usepackage{xcolor}
\usepackage{braket}

\newcommand{\eqref}[1]{(\ref{#1})}
\DeclareMathAlphabet      {\mathitbf}{OML}{cmm}{b}{it}

\newcommand{\tw}{t_\mathrm{w}}
\begin{document}

\title[An experiment-oriented analysis of 2D SG dynamics: A scaling study]{An
  experiment-oriented analysis of 2D spin-glass dynamics: a twelve
  time-decades scaling study}

\author{L.~A.~Fernandez$^{1,2}$, E.~Marinari$^{3,4,5}$, V.~Martin-Mayor$^{1,2}$, G.~Parisi$^{3,4,5}$ and  J.~J.~Ruiz-Lorenzo$^{6,2}$}

\address{$^1$ Departamento de  F\'{\i}sica Te\'orica. Facultad de Ciencias
  F\'{\i}sicas. Universidad Complutense de Madrid. Madrid 28040. Spain.}
\address{$^2$ Instituto de Biocomputaci\'on y
  F\'{\i}sica de Sistemas Complejos (BIFI), 50018 Zaragoza, Spain.}
\address{$^3$ Dipartimento di Fisica, Sapienza
  Universit\`a di Roma, I-00185 Rome, Italy.}
\address{$^4$ Nanotec, Consiglio Nazionale delle Ricerche, I-00185 Rome, Italy.}
\address{$^5$ Istituto Nazionale di Fisica Nucleare, Sezione di Roma 1, I-00185 Rome, Italy.}
\address{$^6$ Departamento de F\'{\i}sica and
  Instituto de Computaci\'on Cient\'{\i}fica Avanzada (ICCAEx), Universidad de
  Extremadura, 06071 Badajoz, Spain.} 
\date{\today}

%%%%%%%%%%%%%%%%%%%%%%%%%%%%%%%%%%%%%%%%%%%%%%%%%%%%%%%%%%%%%%%%%%%%%%
\begin{abstract}
Recent high precision experimental results on spin-glass films ask for
a detailed understanding of the domain-growth dynamics of
two-dimensional spin glasses. To achieve this goal, we numerically
simulate the out-equilibrium dynamics of the Ising spin glass for a
time that spans close to twelve orders of magnitude (from picoseconds
to order of a second), in systems large enough to avoid finite-size
effects. We find that the time-growth of the size of the glassy
domains is excellently described by a single scaling function. A
single time-scale $\tau(T)$ controls the dynamics. $\tau(T)$ diverges
upon approaching the $T=0$ critical point. The divergence of
$\tau(T\to 0)$ is Arrhenius-like, with a barrier height that depends
very mildly on temperature. The growth of this barrier-height is best
described by critical dynamics. As a side product we obtain an
impressive confirmation of universality of the equilibrium behavior of
two-dimensional spin-glasses.
\end{abstract}

\maketitle 

%%%%%%%%%%%%%%%%%%%%%%%%%%%%%%%%%%%%%%%%%%%%%%%%%%%%%%%%%%%%%%%%%%%%%%%
\section{Introduction}

Spin glasses~\cite{mydosh:93,young:98,mezard:86,fisher:91} provide an excellent
model-system to investigate glassy behavior: sluggish glassy dynamics
is observed in a large variety of systems (polymers, supercooled
liquids, colloids, spin glasses, vortex arrays in superconductors,
etc.~\cite{cavagna:09}). Typically, experimental spin-glasses are
studied under out-equilibrium conditions. The disordered experimental
system is quickly cooled from some very high temperature to the
working temperature $T$. As the waiting time $\tw$ increases, the size
of the (glassy) magnetic domains, $\xi(\tw,T)$, grows. A somewhat
indirect experimental procedure can allow to measure $\xi(\tw,T)$
through the Zeeman-effect lowering of free-energy
barriers~\cite{joh:99,marinari:96,guchhait:17}.  Recent numerical
simulations~\cite{janus:08,janus:09,janus:14} have shown that the
$\xi(\tw,T)$ obtained in macroscopic
measurements~\cite{joh:99,guchhait:17} precisely matches the
$\xi(\tw,T)$ obtained in a microscopic computation of the spin-glass
correlation function~\cite{janus:17b}. The Zeeman
method~\cite{joh:99}, however, is not appropriate for precision
measurements of $\xi(\tw,T)$. This is unfortunate, since understanding
in details the temperature and time dependence of the growth of
$\xi(\tw,T)$ is a major issue in the physics of glassy systems.

In this context, an experimental breakthrough has been obtained
recently~\cite{guchhait:14,guchhait:15a,guchhait:15b,guchhait:17,zhai:17}. The
preparation of spin-glass (Cu:Mn) samples of excellent quality in a
thin-film geometry has made it possible to study spin-glasses at the
mesoscale (the film thickness can be varied in the range 9 nm --- 80
nm, while the typical Mn-Mn separation is $5.3$\AA). The film
thickness provides a reference length scale.  For the first time in
the field, \emph{lengths} and \emph{times} are considered on the same
footing in the \emph{same} experiment. This has resulted, for
instance, in an experimental measurement of dimensional crossover from
space dimension $D=3$ to $D=2$ when $\xi(\tw,T)$ grows to the sample
thickness\cite{guchhait:14}.

On the experimental side, $2D$ spin glasses have been analyzed first
in a 1993 paper~\cite{schins:93}. In this case the authors analyzed
data by assuming an activated dynamics.

A pioneering simulation of the $D=2$ Ising spin-glass dynamics could
not resolve whether the dynamics behaves as critical or
activated~\cite{rieger:94} (see also the numerical simulations of
Ref.~\cite{barrat:01}). It is remarkable that after many years the
issue is still open, in spite of recent work~\cite{xu:17,rubin:17}.

It is clear in any case that, when having in mind the behavior of a
film, the theoretical study of a 2D system is only a first step.  The
$D=3$ to $D=2$ crossover can be analyzed in detail from finite-size
(or rather finite-thickness) scaling~\cite{barber:83}. Indeed, after a
block-renormalization of size equal to the film thickness, we are left
with a purely two-dimensional (i.e. single-layer) spin glass system.

Here, we clarify the dynamical behavior of $2D$ spin glasses by means
of a large-scale numerical simulation of the out-equilibrium dynamics
of the $D=2$ Ising spin-glass. The timescale of our simulation spans
close to 12 orders of magnitude (in physical terms, from picoseconds
to close to half a second), thanks to an improvement over the standard
multisite (MUSI) multispin coding technique~\cite{ito:90}, that
dramatically reduces the number of needed random
numbers~\cite{fernandez:15}. In this way, we follow the
microscopic spin-glass coherence length $\xi(\tw,T)$ from virtually
zero to its ultimate equilibrium value $\xi_\mathrm{eq}(T)$. The size
of the simulated systems is large enough to allow a sensible
comparison to experiments. The range of $\xi_\mathrm{eq}(T)$ is large
enough to offer a clear picture of the scaling behavior as temperature
varies. In different dynamical regimes, the dynamics can be classified
as \emph{critical} or \emph{activated}. We provide a quantitative
description of each regime through a scaling function
(Sect.~\ref{sect:scaling}).

This work is organized as follows. In Sect.~\ref{sect:observables} we
describe the model and we define and discuss the basic spin glass
correlation functions (for technical details,
see Appendices A  and B). Our main
result is the scaling analysis presented in
Sect.~\ref{sect:scaling}. Our very accurate data allow us to revisit
the debated issue of universality of the (equilibrium) critical
behavior in Sect.~\ref{sect:universality}. Besides, in
Sect.~\ref{sect:temperature-shift} we investigate a simple
temperature-changing protocol. We present our conclusions in
Sect.~\ref{sect:conclusions}.

%%%%%%%%%%%%%%%%%%%%%%%%%%%%%%%%%%%%%%%%%%%%%%%%%%%%%%%%%%%%%%%%%%%%%%%
\section{The Model and Observables}\label{sect:observables}

\begin{figure}
\includegraphics[width=\columnwidth,angle=0]{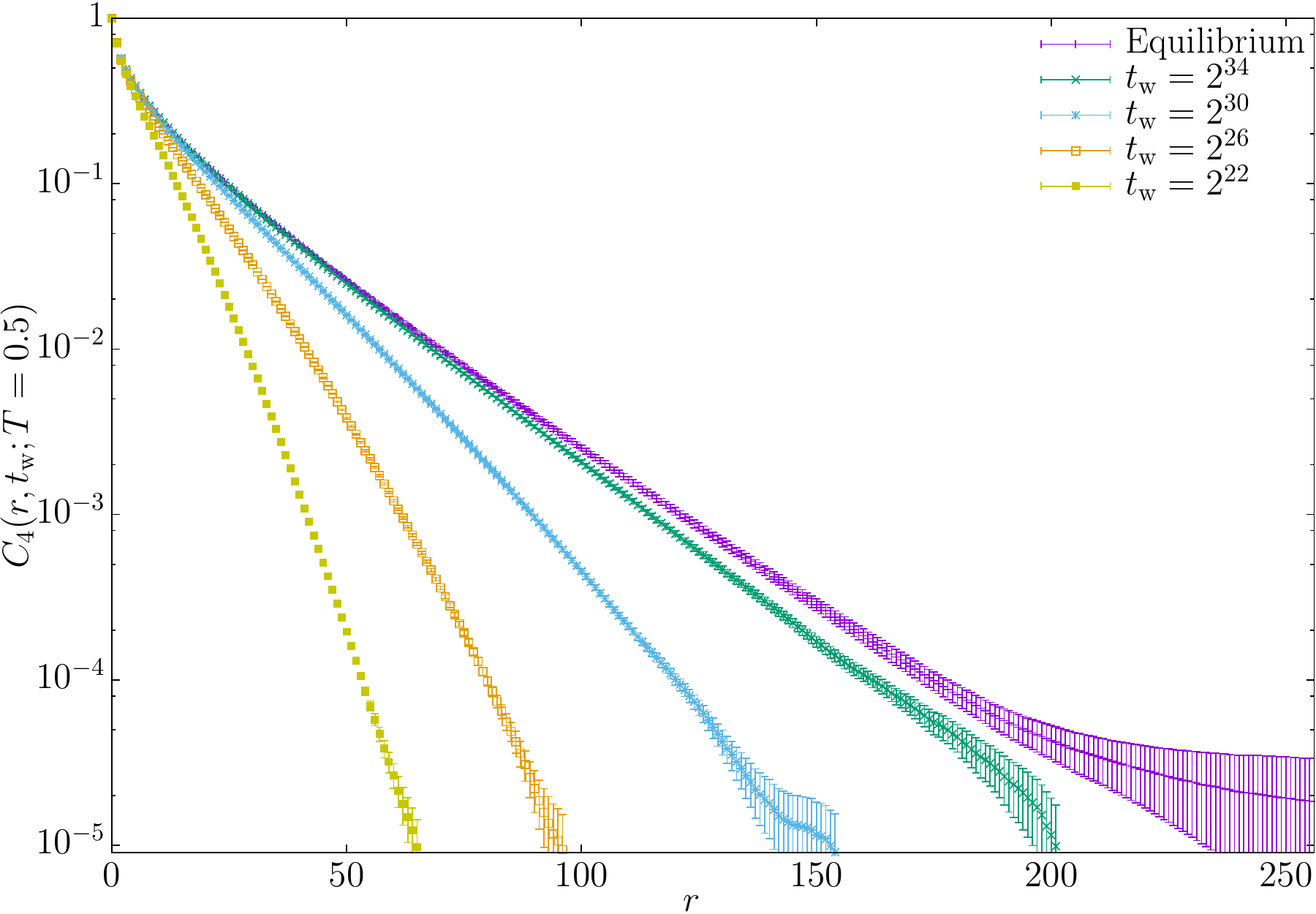}
\caption{The correlation function $C_4(r,\tw)$, Eq.~\eqref{eq:C4-def},
  versus distance $r$ [$\mathitbf{r}=(r,0)$], for several waiting
  times $\tw$ and in the limit of an equilibrated system
  $C^{\mathrm{eq}}_4(r)$ (data for our lowest temperature, $T=0.5$,
  see \ref{sect:simu-details}). The vertical axis represents five
  orders of magnitude. The range of correlations, characterized
  through the $\xi_{12}(\tw)$ coherence length,
  Eq.~\eqref{eq:def-xi-k-kp1} and Fig.~\ref{fig:xioverxieq}, increases
  upon increasing $\tw$ until the equilibrium value
  $\xi_{12}^{\mathrm{eq}}(T)$ is reached [$\xi_{12}^{\mathrm{eq}}(T)$
    diverges at $T=0$, Eq.~\eqref{eq:nu-theta-def}]. While
  $C^{\mathrm{eq}}_4(r)$ decays exponentially in $r$, $C_4(r,\tw)$
  decays super-exponentially
  $C_4(r,\tw)\sim\mathrm{e}^{-(r/{\hat\xi})^\beta}$, with
  $\beta>1$~\cite{fernandez:18b}. Note that the time $C_4(r,\tw)$
  needs to coalesce with $C^{\mathrm{eq}}_4(r)$ depends strongly on $r$.
}
\label{fig:C4-first}
\end{figure}

We have studied the Edwards-Anderson
model~\cite{edwards:75,edwards:76} on a square lattice, with
nearest-neighbors couplings and periodic boundary conditions. Its
Hamiltonian reads
\begin{equation}\label{eq:H}
{\cal H}=-\sum_{\langle\mathitbf{x},\mathitbf{y}\rangle>} J_{\mathitbf{x},\mathitbf{y}} s_\mathitbf{x} s_\mathitbf{y} \,.
\end{equation}
$s_\mathitbf{x}=\pm 1$ are Ising spins.  The couplings
$J_{\mathitbf{x},\mathitbf{y}}=\pm 1$ are chosen independently and
randomly (with 50\% probability). The set of couplings
$\{J_{\mathitbf{x},\mathitbf{y}}\}$, which is chosen at the beginning
of the simulation and kept fixed afterwards, defines a
\emph{sample}. The linear system size, $L=512$ is large enough to be
representative of the thermodynamic limit (see Ref.~\cite{fernandez:18b}).

At the initial time $\tw=0$ we start from random ($T=\infty$)
configurations. The system is suddenly placed at the working temperature $T$,
and from that moment evolves according to
Metropolis dynamics. We measure the time $\tw$
in units of full lattice Metropolis sweeps (one lattice sweep roughly
corresponds to one picosecond~\cite{mydosh:93}).

Our study is based on the analysis of the overlap-overlap correlation function
(see Ref.~\cite{janus:09b} for a detailed discussion)
\begin{equation}\label{eq:C4-def}
C_4(\mathitbf{r};\tw) =
E(q^{a,b}(\mathitbf{x},\tw) q^{a,b}(\mathitbf{x}+\mathitbf{r},\tw))\,,
\end{equation}
built from the replica overlaps
\begin{equation}
q^{a,b}(\mathitbf{x},\tw) = s^{(a)} (\mathitbf{x},\tw)
s^{(b)}(\mathitbf{x},\tw) \,.
\end{equation}
In Eq.~\eqref{eq:C4-def}, we have denoted by $E(\cdots)$ the
average over the couplings, the thermal noise and the random initial
conditions. The $\{s^{(a)} (\mathitbf{x},\tw)\}$ are \emph{real
  replicas} ($a$ is the  replica index): different replicas
evolve under the same set of couplings
$\{J_{\mathitbf{x},\mathitbf{y}}\}$ but are otherwise statistically
independent. 

Following Refs.~\cite{janus:09b,janus:08b}, we consider
displacement vectors along one of the lattice axis, either
$\mathitbf{r}=(r,0)$ or $\mathitbf{r}=(0,r)$, and use the shorthand
$C_4(r;\tw)$. The $r$ and $\tw$ dependencies of $C_4(r;\tw)$ are
shown in Fig.~\ref{fig:C4-first}. We compute the coherence
length (the typical size of the glassy domains)
\begin{equation}\label{eq:def-xi-k-kp1}
  \xi_{k,k+1}(\tw)\equiv I_{k+1}(\tw)/I_k(\tw)\,,
\end{equation}
by means of the integrals
\begin{equation}\label{eq:Ik-def}
  I_k(\tw)\equiv\int_0^{\infty}\mathrm{d}\,r\ r^k C_4(r;\tw)\,.
\end{equation}
Following recent work~\cite{janus:17b,janus:09b,janus:08b,janus:16},
we focus our attention on the $k=1$ estimate  $\xi_{12}(\tw)$.
Further details on the computation of the integrals in
Eq.~\eqref{eq:Ik-def} are given in \ref{sect:truncation}.

Eventually, we have been able to equilibrate the
system.\footnote{Strictly speaking, an infinite system never fully
  equilibrates. One could rather think of an~\emph{equilibration
    wave-front}: $C_4(r,\tw)\approx C_4^{\mathrm{eq}}(r)$ if, at time
  $\tw$, $r$ lies behind the wave-front (see
  Fig.~\ref{fig:C4-first} and Ref.~\cite{fernandez:18b}).  Once the
  $C_4(r,\tw)$ equilibrates up to a distance (say) $r=6\,
  \xi_{12}^{\mathrm{eq}}(T)$, we can regard the system as
  equilibrated for all practical purposes.} In this limit we define
\begin{equation}
\xi_{12}^{\mathrm{eq}}(T)=\lim_{\tw\to\infty} \xi_{12}(\tw,T)\,.
\end{equation}
In our simulations, $\xi_{12}^{\mathrm{eq}}(T)$ ranges from
$\xi_{12}^{\mathrm{eq}}(T=1.1)\approx 4.3$ to $\xi_{12}^{\mathrm{eq}}(T=0.5)\approx 39.4$:
this is why we expect that $L=512$ is large enough to accommodate $L\to\infty$
conditions~\cite{janus:08b,janus:16} [we also check that, within our small
statistical errors, $C_4(r=L/2,\tw)=0$]. In fact, if one takes first the limit
$L\to\infty$ and only afterwards goes to low $T$, we expect
\begin{equation}\label{eq:nu-theta-def}
0<\, \lim_{T\to 0}\, [T^{\nu} \, \xi_{12}^{\mathrm{eq}}(T)]\, < +\infty\,,\ 1/\nu=-\theta\,.
\end{equation}
Recently, the stiffness exponent $\theta$ has been computed in an
impressive $T=0$ simulation for Gaussian couplings, with the result
$\theta=-0.2793(3)$~\cite{khoshbakht:17}(one expects $\theta= −1/\nu$, which was confirmed in
previous Gaussian couplings simulations, see
e.g. Refs.~\cite{katzgraber:04,fernandez:16b}). We show in
Sect.~\ref{sect:universality} that this description holds as well for
our case of $J=\pm 1$ couplings.

%%%%%%%%%%%%%%%%%%%%%%%%%%%%%%%%%%%%%%%%%%%%%%%%%%%%%%%%%%%%%%%%%%%%%%%
\section{Scaling properties of the correlation length.}\label{sect:scaling}

\begin{figure}
\includegraphics[width=\columnwidth,angle=0]{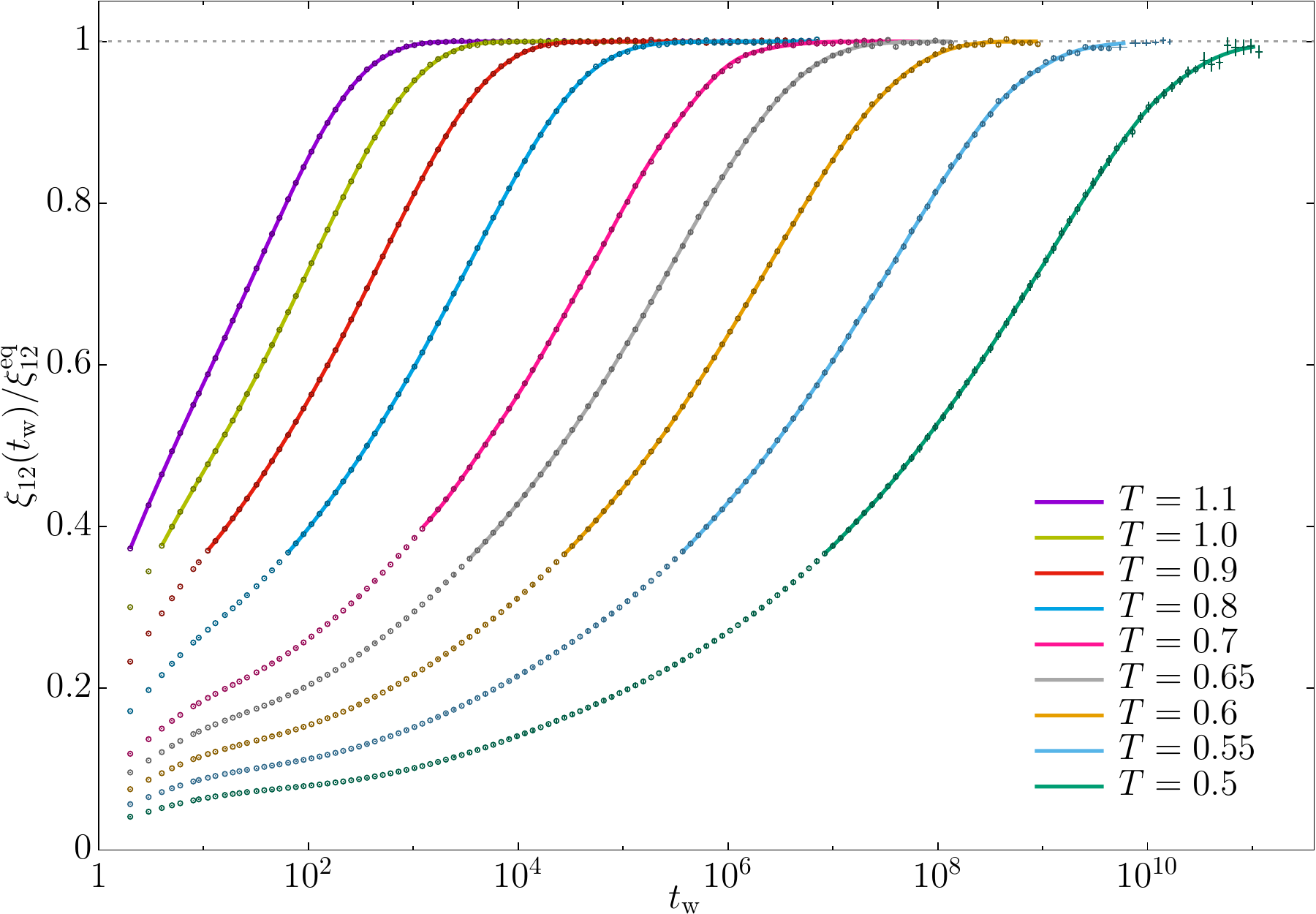}
\caption{Ratio of characteristic sizes
  $\xi_{12}(\tw,T)/\xi_{12}^{\mathrm{eq}}(T)$ as a function of
  simulation time. Most data shown are for $N_R=256$ replicas
  ($T>0.5$) or $N_R=264$ replicas ($T=0.5$). However, the values of
  $\xi_{12}^{\mathrm{eq}}(T=0.55)$ and
  $\xi_{12}^{\mathrm{eq}}(T=0.5)$ were obtained from much longer
  simulations, with a smaller number of replicas
  (see \ref{sect:simu-details}).
  Continuous lines are fits to Eq.~\eqref{eq:fit-xioverxieq}.}
\label{fig:xioverxieq}
\end{figure}

Our analysis will be based on the quotient
$\xi_{12}(\tw,T)/\xi_{12}^{\mathrm{eq}}(T)$, which is shown in
Fig.~\ref{fig:xioverxieq}. 

We have interpreted the data in figure \ref{fig:xioverxieq} through a single
scaling function
\begin{equation}\label{eq:scaling-F}
  \frac{\xi_{12}(\tw,T)}{\xi_{12}^\mathrm{eq}(T)}
  ={\cal F}\left(\frac{t_w}{\tau(T)}\right)\
  +\ {\cal O}\Big(\, [\xi_{12}(\tw,T)]^{-\omega},
  [\xi_{12}^{\mathrm{eq}}(T)]^{-\omega}\,\Big)\,.
\end{equation}
In the above expression $\omega$ is some sort of unknown
corrections-to-scaling exponent. When writing
Eq.~\eqref{eq:scaling-F}, it is obvious that we have some hope that
the scaling function ${\cal F}(x)$ will be universal (the same type
of scaling, but with a different
scaling function, appears for the Langevin dynamics of the scalar
free-field, see Ref.~\cite{fernandez:18b}).\footnote{
Here, we consider only times  and temperatures such that $\xi_{12}(\tw,T)/ \xi_{12}^\mathrm{eq}(T)$ remains fixed in the limit $T\to 0$, where $\xi_{12}^\mathrm{eq}(T)$ diverges.} 

We shall first proceed to a qualitative discussion of our scaling hypothesis
(see Section~\ref{subsect:qualitative}), deferring the more detailed study to
Section~\ref{subsect:quantitative}.

\subsection{Qualitative analysis of Eq.~\eqref{eq:scaling-F}}\label{subsect:qualitative}

\begin{figure}
\includegraphics[width=\columnwidth,angle=0]{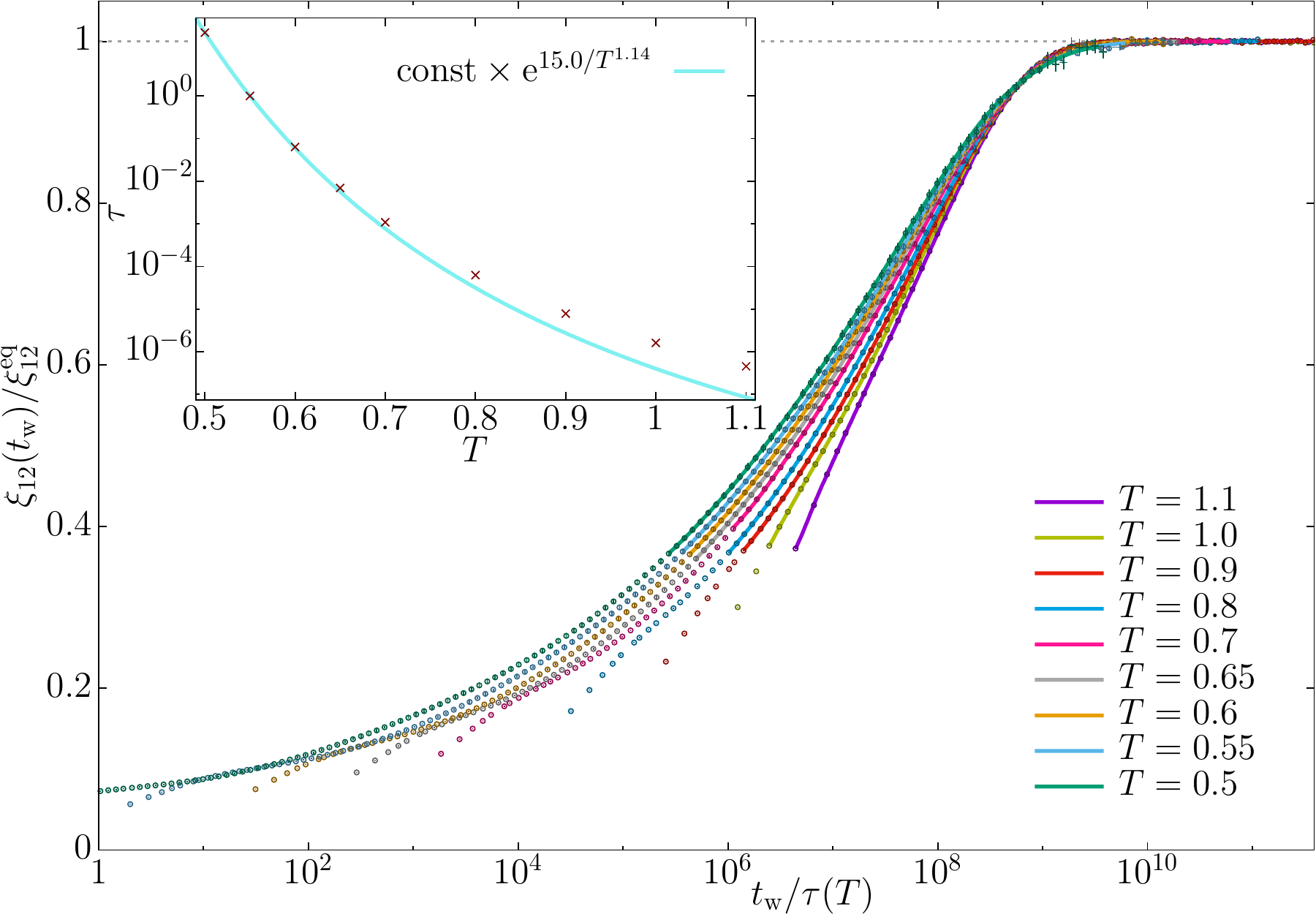}
\caption{Data for $\xi_{12}(\tw,T)/\xi_{12}^{\mathrm{eq}}(T)$ from
  Fig.~\ref{fig:xioverxieq} with a temperature-dependent time
  rescaling factor $\tau(T)$, as suggested by
  Eq.~\eqref{eq:scaling-F}. In order to chose $\tau(T)$ one needs
  (somewhat arbitrary) normalization conditions. Our choices have been
  (i) $\tau(T=0.55)=1$ and (ii) the curves for all temperatures must
  cross at the value of $\tw/\tau(T)$ such that
  $0.95=\xi_{12}/\xi_{12}^{\mathrm{eq}}$. {\bf Inset:} The
  time-rescaling factors $\tau(T)$ from the main panel as a function
  of temperature. The curve shows that the $\tau(T)$ at low
  temperatures are nicely described by a slightly modified Arrhenius
  law.}
\label{fig:xioverxieq_rescaled}
\end{figure}

As we anticipated, see Fig.~\ref{fig:xioverxieq_rescaled}, we can find
time-rescaling factors $\tau(T)$ such that the numerical data approach
as much as possible the functional form anticipated in
Eq.~\eqref{eq:scaling-F}. An scaling curve is approached when
temperature decreases.  Nevertheless, scaling corrections are clearly
visible, given the high accuracy of our numerical data. These scaling
corrections decrease as $\xi_{12}^{\mathrm{eq}}(T)$
increases.\footnote{Let us recall from Ref.~\cite{fernandez:18b} the
  equilibrium values for $\xi_{12}^{\mathrm{eq}}(T):
  \xi_{12}^{\mathrm{eq}}(1.1)=4.272(6)\,,\xi_{12}^{\mathrm{eq}}(1.0)=5.332(10)\,,
  \xi_{12}^{\mathrm{eq}}(0.9)=6.904(13)\,,
  \xi_{12}^{\mathrm{eq}}(0.8)=9.387(27)\,,
  \xi_{12}^{\mathrm{eq}}(0.7)=13.585(63)\,,
  \xi_{12}^{\mathrm{eq}}(0.65)=16.845(95)\,,
  \xi_{12}^{\mathrm{eq}}(0.6)=21.50(13)\,,
  \xi_{12}^{\mathrm{eq}}(0.55)=28.52(25)$ and
  $\xi_{12}^{\mathrm{eq}}(0.5)=39.36(47)$} We note as well that, for
any given temperature, the smaller $\xi_{12}(\tw,T)$ the larger become
the corrections to scaling.

As for the time-rescaling factor $\tau(T)$, see
Fig.~\ref{fig:xioverxieq_rescaled}--inset, it can be described
(particularly at low temperatures), by a modified Arrhenius law in
which the exponent of $1/T$ is slightly above 1. We shall come back to this effect below, see Eq.~\eqref{eq:barreras-criticas}.

\subsection{Quantitative analysis of Eq.~\eqref{eq:scaling-F}}\label{subsect:quantitative}
In order to be quantitative, we have made two checks on
Eq.~\eqref{eq:scaling-F}. First we need to introduce some notations. Let
$\tw(f;T)$ be the time needed to reach a fraction $f$ of the
equilibrium correlation length
\begin{equation}\label{eq:f-def}
f=\frac{\xi_{12}(\tw(f;T),T)}{\xi_{12}^\mathrm{eq}(T)}\,.
\end{equation}
Now, in order to compute $\tw(f;T)$, we need an interpolating scheme
delivering $\xi_{12}(\tw)/\xi_{12}^{\mathrm{eq}}(T)$ as a continuous
function of $\tw$ [obviously we compute $\xi_{12}(\tw)$ for discrete
  values of $\tw$ only]. In order to address this problem, we shall
need to make first some general considerations.

In principle, the approach to equilibrium at temperature $T$ for any
physical quantity is ruled by the same set of characteristic
autocorrelation times
$\tau_1(T)>\tau_2(T)>\tau_3(T)>\ldots$~\cite{sokal:97}. In a system of
$L^2$ spins, the number of characterstic times is $2^{L^2}-1$. So, when
the starting state at $\tw=$ is fully disordered, as it is our case,
the time evolution of any particular quantity, $A$, behaves as
\begin{equation}\label{eq:A-tw-T}
A(\tw,T)=A^\mathrm{eq}(T)+\sum_{\alpha=1}^{2^{L^2}-1} a_\alpha(T)\mathrm{e}^{-\tw/\tau_\alpha(T)}\,.
\end{equation} 
In the above expression, the amplitudes $a_\alpha(T)$ are
magnitude-specific, but the times $\tau_\alpha(T)$ are the same for
all quantities. Now, when $L$ becomes large, the discrete set of times
$\tau_\alpha$ becomes a continuous distribution. Let us specialize to $\xi_{12}(\tw,T)$, which relates directly to the experimental non-linear response~\cite{janus:17b}. In the limit of large $L$,  Eq.~\eqref{eq:A-tw-T} takes the form of a Laplace-like decomposition (see e.g. Ref.~\cite{ogielski:85}):
\begin{equation}\label{eq:Laplace-ideal}
\frac{\xi_{12}(\tw,T)}{\xi_{12}^{\mathrm{eq}}(T)}=1-\int_{\mathrm{log}\,\tau_{\mathrm{min}}(T)}^{\mathrm{log}\,\tau_{\mathrm{max}}(T)}\mathrm{d}\,(\mathrm{log}\,\tau)\ \rho_\xi(\log\, \tau,T)\,\mathrm{e}^{-\tw/\tau}\,.
\end{equation}
In the above expression, the time-distribution $\rho_\xi(\log\,
\tau,T)$ is specific to $\xi$ [but the support of the distribution,
  namely the $\tau$ in which $\rho_\xi(\log\, \tau,T)>0$, would be the
  same for any other quantity, recall
  Eq.~\eqref{eq:A-tw-T}]. Furthermore, from $\xi(\tw=0,T)=0$ we get
\begin{equation}
1=\int_{\mathrm{log}\,\tau_{\mathrm{min}}(T)}^{\mathrm{log}\,\tau_{\mathrm{max}}(T)}\mathrm{d}\,(\mathrm{log}\,\tau)\ \rho_\xi(\log\, \tau,T)\,.
\end{equation}
Unfortunately, obtaining the distribution of characteristic times
$\rho_\xi(\log\, \tau,T)$, with support in the interval $\log\tau_\mathrm{min}(T)<\log\tau<\log\tau_\mathrm{max}(T)$, through a numerical inversion of its Laplace
transform, 1- $\xi_{12}(\tw)/\xi_{12}^{\mathrm{eq}}(T)$, is an
ill-posed mathematical problem (see for example Ref.~\cite{epstein:08}).
Hence, in order to make progress, we discretize Eq.~\eqref{eq:Laplace-ideal}
by making the strong
assumption of a very smooth $\rho_\xi(\log\, \tau,T)$. In other words we
assume $\rho_\xi$ to be such
that it can be faithfully interpolated by its value at a very small
number of  points at constant logarithmic distance, $\tau_n=\tau/b^n$,
$n=0,1,\ldots,R-1$:
\begin{equation}\label{eq:fit-xioverxieq}
\frac{\xi_{12}(\tw,T)}{\xi_{12}^{\mathrm{eq}}(T)}=
1\ -\ \sum_{n=0}^{R-1}\, c_n \,\mathrm{e}^{-\frac{\tw}{\tau_n}}\ +\ \sum_{n=0}^{R-2}\frac{c_n+c_{n+1}}{2}\,\mathrm{e}^{-\frac{\tw\sqrt{b}}{\tau_n}}\,.
\end{equation}
Our rationale for including the second sum in
Eq.~\eqref{eq:fit-xioverxieq} is that, while $\rho_\xi(\log\, \tau,T)$
is a slowly varying function of $\mathrm{log}\,\tau$, certainly
$\mathrm{e}^{-\tw/\tau}$ has a strong dependency in
$\mathrm{log}\,\tau$. This strong variation makes it advisable to
interpolate $\mathrm{e}^{-\tw/\tau}$ between its values at $\tau=\tau_n$
and $\tau=\tau_{n+1}$.

Focusing our attention on fractions
$f>0.36$ (see Eq.~\eqref{eq:f-def} and Fig.~\ref{fig:xioverxieq})
we have found excellent (and very stable) fits to
Eq.~\eqref{eq:fit-xioverxieq} with $R$ as small as $7$ (but for $T=1.1$ and
$T=0.5$, where $R=6$). The fitting parameters were the maximum
time $\tau$, the logarithmic $\tau$-spacing $\mathrm{\log}\,b$, and
the amplitudes $c_n$'s. Due to the strong statistical correlation for
the different $\tw$'s, we employ the jackknife as implemented
in~\cite{yllanes:11}: we fit for each jack-knife block (using for all
blocks the diagonal covariance matrix), and compute errors from the
blocks fluctuations.

The reason for disregarding small-$f$ fractions in our analysis is the
 two-steps mechanism that governs the behavior of $\xi_{12}(\tw)$ for
the model with discrete couplings $J=\pm 1$. Indeed, with $J=\pm 1$ we
have an energy-gap $\Delta E=4$ between the ground state and the first
excited state~\cite{katzgraber:05}, which causes a peculiar  short-time behavior at low
temperatures.  In the first relaxation step, $\xi_{12}(\tw)$ reaches
very quickly a plateau at $\xi_{12}\approx 2$.  This plateau is
visible in Fig.~\ref{fig:xioverxieq}, although its height apparently
decreases upon lowering $T$, due to the normalization with
$\xi_{12}^\mathrm{eq}(T)$.  Only after a time
$\tw\sim\mathrm{e}^{4/T}$ the relaxation proceeds, and $\xi_{12}(\tw)$
grows significantly. From the point of view of
Eq.~\eqref{eq:scaling-F}, the plateau causes uninteresting
scaling-corrections of order $\sim 2/\xi_{12}^\mathrm{eq}(T)$. These corrections 
make it advisable to avoid the small-$f$ region, defining  the safe range $4.3
\lesssim \xi_{12}^\mathrm{eq}(T)\lesssim 39.4$ in our simulations.

\begin{figure}
\includegraphics[width=\columnwidth,angle=0]{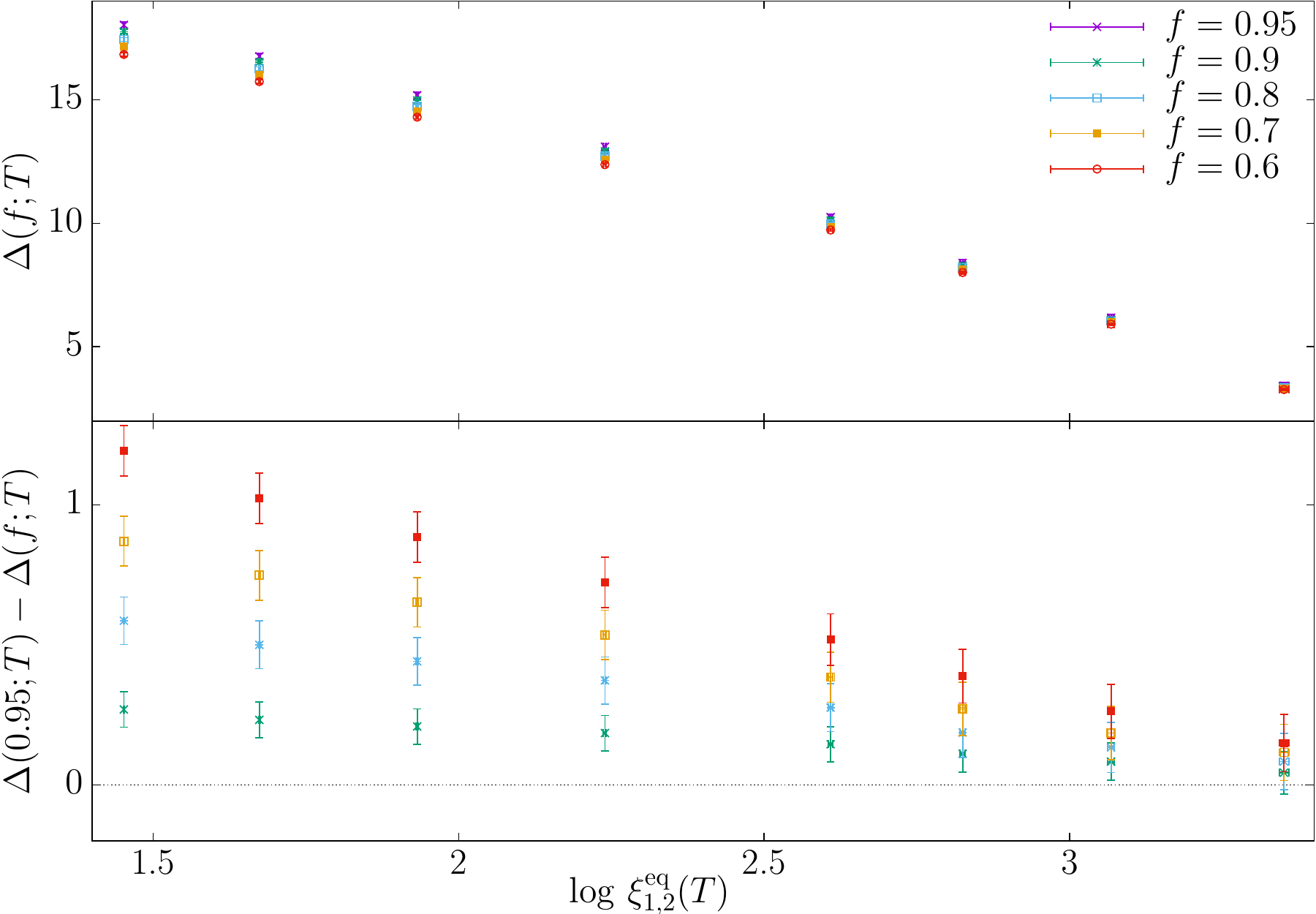}
\caption{{\bf Top:} Should the scaling~\eqref{eq:scaling-F} be exact,
  the logarithmic-time difference $\Delta(f;T)$ defined in
  Eq.~\eqref{eq:check-1} would be $f$-independent at all temperatures.
  Indeed, the $T$-dependence in $\Delta(f;T)$ is much larger than a
  tiny (but visible at high $T$) $f$-dependence. {\bf Bottom:} taking
  the difference $\Delta(0.95;T)-\Delta(f;T)$ eliminates most of the
  $T$-dependence in the top-panel data. We see that the
  $f$-dependence of $\Delta(f;T)$ quickly decreases (indeed
  $\Delta(f;T)-\Delta(0.95;T)$ approaches zero) as
  $\xi_{12}^{\mathrm{eq}}(T)$ grows.}
\label{fig:Check-1}
\end{figure}

\begin{figure}
\includegraphics[width=\columnwidth,angle=0]{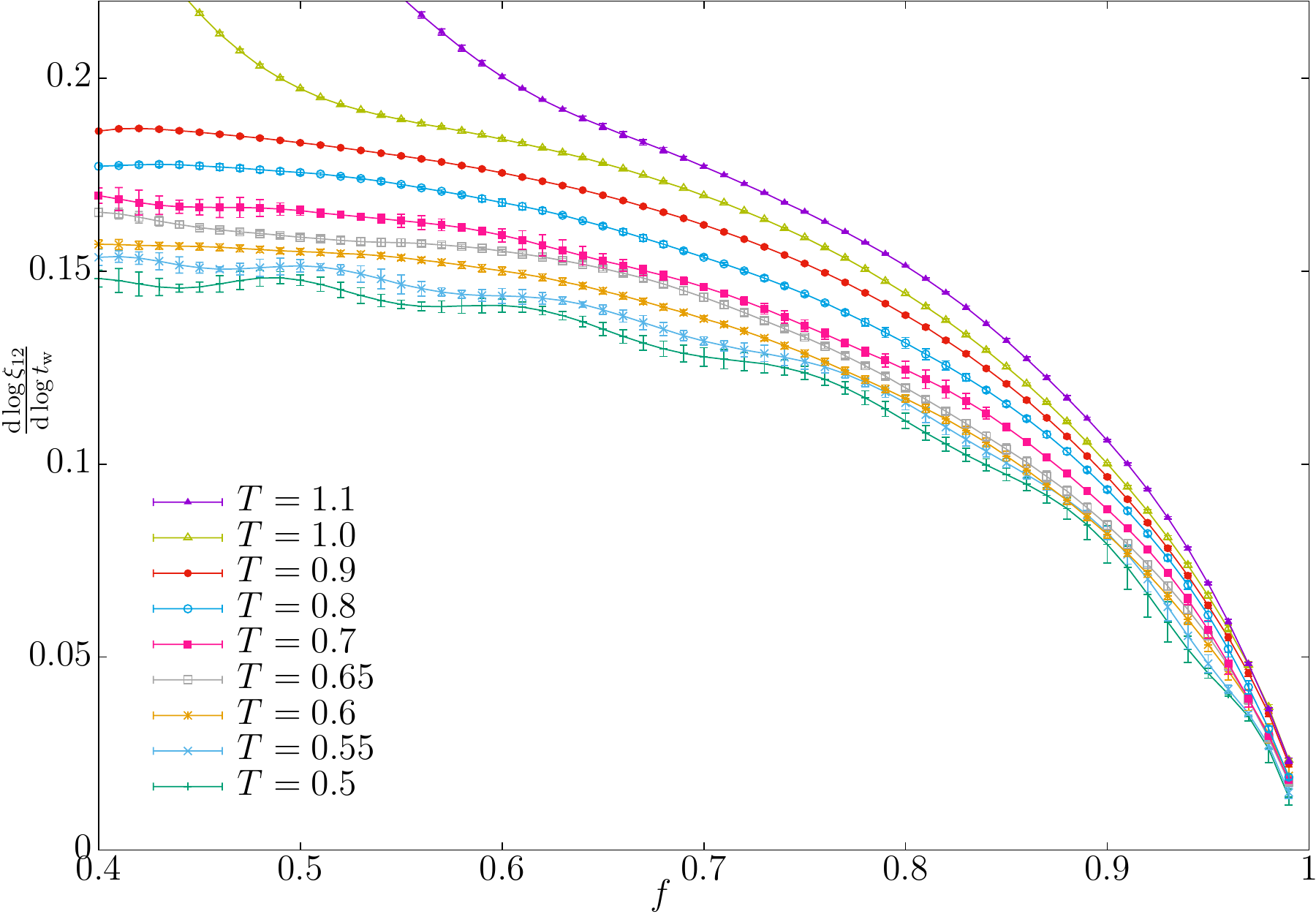}
\caption{Numerical illustration of Eq.~\eqref{eq:derivative-check}.}
\label{fig:Check-2}
\end{figure}

At this point, counting on the continuous time
interpolation~\eqref{eq:fit-xioverxieq}, we can
present our two checks on Eq.~\eqref{eq:scaling-F}.
\begin{itemize}
\item 
Let us fix a reference temperature $T^*$ (it will be $T^*=0.5$ in our case).
If we disregard correction to scaling, Eq.~\eqref{eq:scaling-F} implies
that
\begin{equation}\label{eq:check-1}
\Delta(f;T)\equiv \log {\tw(f;T^*)}\ -\ \log {\tw(f;T)}\,,  
\end{equation}
is independent of $f$ for any fixed temperature. As figure
\ref{fig:Check-1} shows, this seems to be truly the case for large
$\xi_\mathrm{eq}$ but not quite so at larger temperatures (this is very
reasonable: we only expect an
universal behavior to appear in the scaling limit
$\xi_\mathrm{eq}\to\infty$).

\item In the limit of large correlation-length, where corrections to
  scaling can be neglected, one should have that
\begin{equation}\label{eq:derivative-check}
  \left. \frac{\mathrm{d} \log [\xi(\tw)]}{\mathrm{d} \log \tw}
  \right|_{\tw=\tw(f;T)}=
  \left.\frac{\mathrm{d} \log {\cal F}(x)}{\mathrm{d} \log x}
  \right|_{x={\cal F}^{-1}(f)}\,,
\end{equation}
[we estimated the derivatives from the interpolating
  function~\eqref{eq:fit-xioverxieq}].  Fortunately, see figure
\ref{fig:Check-2}, our data for the l.h.s. of
Eq.~\eqref{eq:derivative-check} reach the limit of
large-$\xi_{12}^\mathrm{eq}(T)$ rather fast. Indeed, we have
extrapolated linearly in $1/\xi_{12}^\mathrm{eq}(T)$ the derivatives
at $f=0.4$. Including in the fit all points with
$\xi_{12}^\mathrm{eq}(T)>10$, which corresponds to $T\leq 0.7$, the
limiting derivative turns out to be $0.136(3)$ [the derivative at
$T=0.5$ and $f=0.4$ is $0.148(2)$].
\end{itemize}
An added bonus from (\ref{eq:derivative-check}) and Fig.
\ref{fig:Check-2}, is the scaling
\begin{equation}\label{eq:hat-z-def}
{\cal F}(x\to 0)\propto x^{1/\hat z}\,,\ \hat z\approx 7\,.
\end{equation}
Therefore, we have obtained that, when $1\ll \xi_{12}(\tw)$
(so that corrections to
scaling can be neglected), but still $\xi_{12}(\tw)\ll \xi_{12}^{\mathrm{eq}}$ (so
we are still far from equilibrium), one should observe
\begin{equation}
  \xi_{12}(\tw;T) \propto \xi_{12}^\mathrm{eq}(T)
  \left(\frac{\tw}{\tau(T)}\right)^{1/\hat z}\,.
\end{equation}

\begin{figure}
\includegraphics[width=\columnwidth,angle=0]{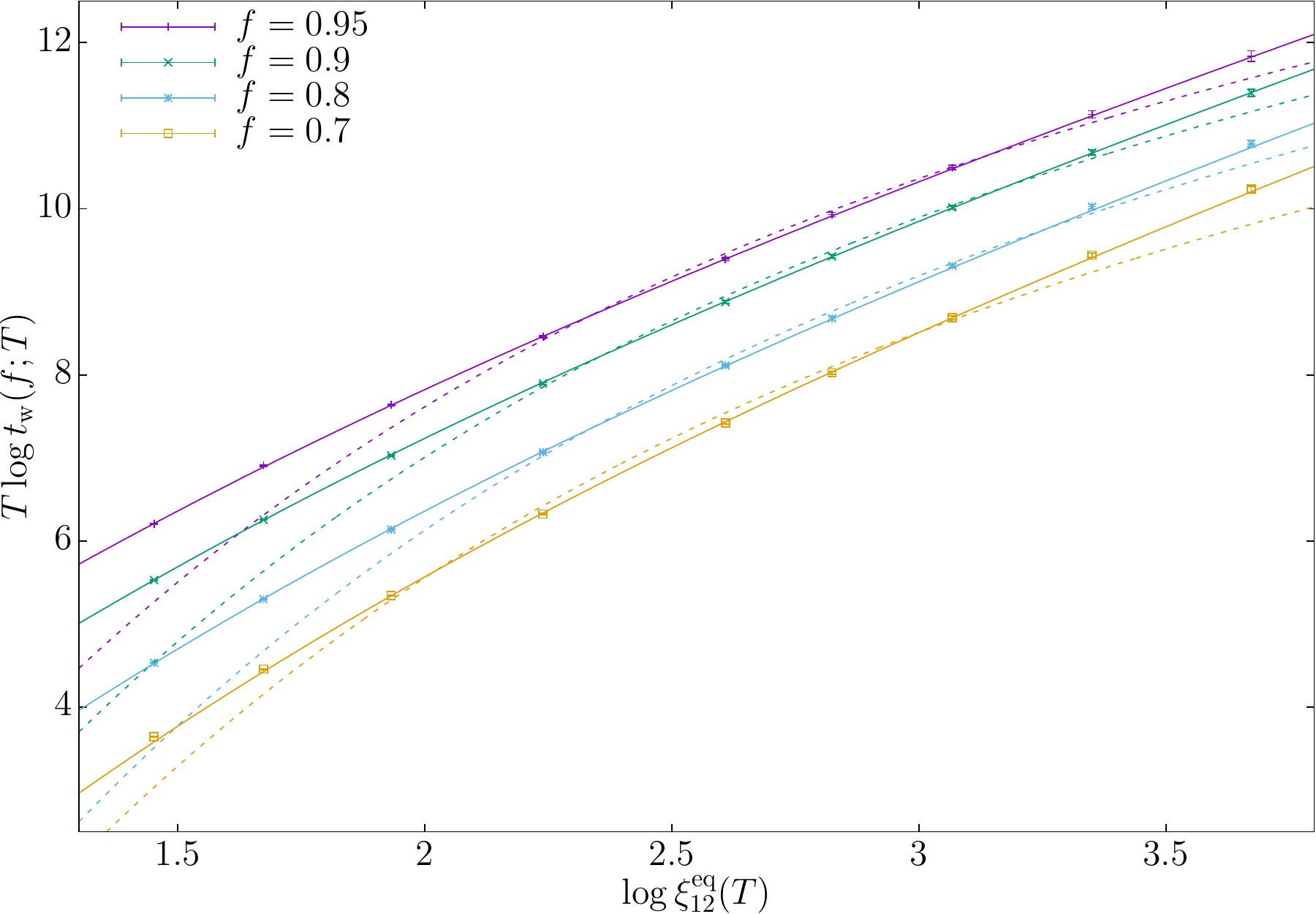}
\caption{$T\log \tw(f;T)$, as a function of
  $\log\xi_{12}^\mathrm{eq}(T)$. The dashed-lines are fits to a fixed
  barrier-height description, Eq.~\eqref{eq:Arrhenius}, where data for
  $T>0.8$ were not used in the fit.  Note that, because of the
  sizable temperature-dependence in our problem, $T\log \tw(f;T)$
  cannot be regarded as an effective barrier height. The continuous
  lines are fits to a mildly diverging barrier-height description,
  Eq.~\eqref{eq:Arrhenius-modified}. The exponent $z = 1.71(5)$ was
  obtained from the fit for $f = 0.95$. All other fits to
  Eq.~\eqref{eq:Arrhenius-modified} use this value of $z$ and have an
  acceptable $\chi^2/\mathrm{dof}$ (in the case of $f = 0.7$ we had to
  discard the $T = 1.1$, $1.0$ data).}
\label{fig:Arrhenius}
\end{figure}

\subsection{The time scale $\tau(T)$}
Equation ~\eqref{eq:scaling-F} tells us that, barring corrections to scaling,
\begin{equation}
\tw(f;T)= \tau(T) {\cal F}^{-1}(f)\,.
\end{equation}
Anticipating some, maybe modified, Arrhenius behavior ($\tau(T)\propto \mathrm{e}^{B/T}$), we observe that
\begin{equation}
T\log \tw(f;T)= T \log \tau(T)\ +\ T\log {\cal F}^{-1}(f)\,.
\end{equation}
A fixed barrier-height, purely Arrhenius scaling would be
\begin{equation}\label{eq:Arrhenius}
T\log \tw(f;T)= B + T a(f)\,,
\end{equation}
where both the barrier height $B$ and the amplitude $a(f)$ are $T$
independent.  As the dashed lines in figure \ref{fig:Arrhenius} show,
this fixed-barrier description is not quantitatively accurate, but it
is a good first approximation. Hence, we have tried a small modification with a
mildly diverging barrier-height
\begin{equation}\label{eq:Arrhenius-modified}
T\log \tw(f;T)= B\ +\ z\log \xi_{12}^\mathrm{eq}(T)\ +\ T a(f)\,.
\end{equation}
The quality of these description is as good as one could hope (we find
$z = 1.71(5)$, see the full lines in Fig.~\ref{fig:Arrhenius}). Note
that this equation implies that
\begin{equation}\label{eq:barreras-criticas}
\tau(T) \propto \mathrm{e}^{B/T} [\xi_{12}^\mathrm{eq}(T)]^{z/T}\,.
\end{equation}
The alert reader will recall
Fig.~\ref{fig:xioverxieq_rescaled}--inset.  Indeed, combining
Eqs.~\eqref{eq:barreras-criticas} and~\eqref{eq:nu-theta-def}, we find
for small temperatures $\log\tau(T)\sim
\frac{1}{T}(B+z\nu\log\frac{T^*}{T})$, where $(T^*)^\nu$ is a
scaling amplitude. This behaviour is numerically indistinguishable form
the one encountered in Fig.~\ref{fig:xioverxieq_rescaled}--inset,
namely $\log\tau(T) \sim \frac{B}{T^{1+\epsilon}}$ where $\epsilon$ is
a small quantity.

Let us finally mention that we have also tried a power-law fit (not shown):
\begin{equation}\label{eq:Arrhenius-modified-power}
T\log \tw(f;T)= c(f)\xi^\Psi_\mathrm{eq}(T)\ +\ T a(f)\,.
\end{equation}
The quality of this second fit is comparable to the one of
Eq.~\eqref{eq:Arrhenius-modified}. The resulting exponent is $\Psi=0.121(3)$,
small enough to mimic a logarithmic behavior.

\begin{figure}
\includegraphics[width=\columnwidth,angle=0]{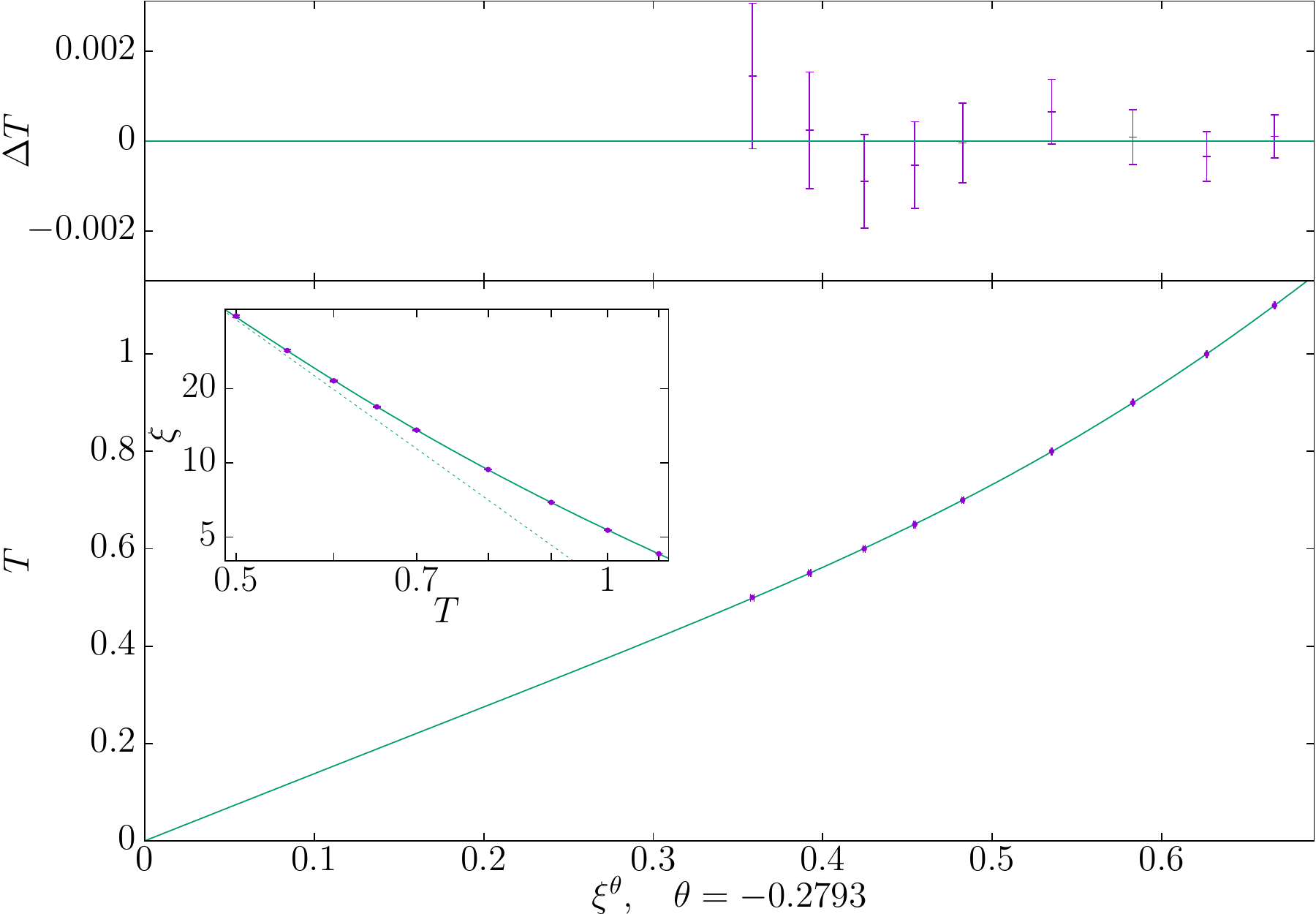}
\caption{{\bf Main panel}: Temperature as a function of the
  equilibrium correlation-length, to the power $\theta$ (the value of
  $\theta$ was taken from Ref.~\cite{khoshbakht:17}). The continuous
  line is a fit to Eq.~\eqref{eq:T_xi} (we truncated the series by
  keeping only terms with $k\leq 3$). The quality of the fit is
  quantified by its figure of merit $\chi^2/\mathrm{dof}=3.1/5$. {\bf
    Inset:} Equilibrium correlation-length as a function of
  temperature (same data from main panel), in logarithmic scale. The
  continuous line is the inverse function as computed for the fit
  shown in the main panel. The dashed line is the inverse function as
  computed from the fit's leading term, namely $a_0
  [\xi_{12}^\mathrm{eq}]^\theta$. The difference betwen the two lines
  quantifies the effect of the corrections-to-scaling terms in
  Eq.~\eqref{eq:T_xi}. {\bf Top}: deviates from fit $\Delta
  T=T-T(\xi_{12}^{\mathrm{eq}})$ as a function of
  $[\xi_{12}^{\mathrm{eq}}]^\theta$. The errors reported are $\delta
  T=[T(\xi_{12}^{\mathrm{eq}}-\delta\xi_{12}^{\mathrm{eq}})-T(\xi_{12}^{\mathrm{eq}}+\delta\xi_{12}^{\mathrm{eq}})]/2$,
  where $\delta\xi_{12}^{\mathrm{eq}}$ is the statistical error for
  $\xi_{12}^{\mathrm{eq}}$.}\label{fig:T_xi}
\end{figure}

%%%%%%%%%%%%%%%%%%%%%%%%%%%%%%%%%%%%%%%%%%%%%%%%%%%%%%%%%%%%%%%%%%%%%%%
\section{On Universality.}\label{sect:universality}

The issue of universality in $2D$ Ising spin-glasses is greatly
complicated by the fact that the critical temperature is $T=0$. At
exactly $T=0$, several renormalization group fixed points are
relevant~\cite{amoruso:03}. However, for an infinite system and $T>0$,
one of those fixed-points dominates (the one corresponding to
Gaussian-distributed couplings $J$), implying a single universality
class~\cite{jorg:06b,parisen:10,thomas:11,parisen:11,jorg:12,fernandez:16b}
(yet, see Refs.~\cite{lundow:15a,lundow:17} for a dissenting view).

However, evidence for universality is at present much weaker for the
thermal critical exponent $\nu$. Indeed in ref. \cite{fernandez:16b},
we could not study the $\nu$ critical exponent in equilibrium for the
$J=\pm 1$ model. The reason for this failure was that, on systems of
finite size, different RG fixed points exchange dominance upon
lowering the temperature for systems of a fixed size. Here, we have
two major advantages: (i) the limit $L\to\infty$ has been safely taken
and (ii) we have at our disposal a beautiful recent determination of
the stiffness exponent $\theta=-0.2793(3)$~\cite{khoshbakht:17} (one
expects $\theta=-1/\nu$). We are explicitly assuming universality, as
we are taking $\theta$ from a Gaussian-couplings computation, and
applying it to our $J=\pm 1$ data.

Following now Ref.~\cite{fernandez:16b}, we invert the function
$\xi_{12}^\mathrm{eq}(T)$ and study the temperature as a function of the
equilibrium correlation-length. The non-linearity of the scaling fields
implies that
\begin{equation}\label{eq:T_xi}
  T(\xi_{12}^\mathrm{eq})= a_0 [\xi_{12}^\mathrm{eq}]^\theta
  (1+ \sum_{k=1}^{\infty} a_k [\xi_{12}^\mathrm{eq}]^{2k\theta} )\,.
\end{equation}
Fig.~\ref{fig:T_xi} shows that indeed this description is
quantitatively very accurate, providing additional strong support
to universality.

%%%%%%%%%%%%%%%%%%%%%%%%%%%%%%%%%%%%%%%%%%%%%%%%%%%%%%%%%%%%%%%%%%%%%%%
\section{Analysis of the Temperature Shift}\label{sect:temperature-shift}

\begin{figure}
\begin{center}
\includegraphics[width=\columnwidth,angle=0]{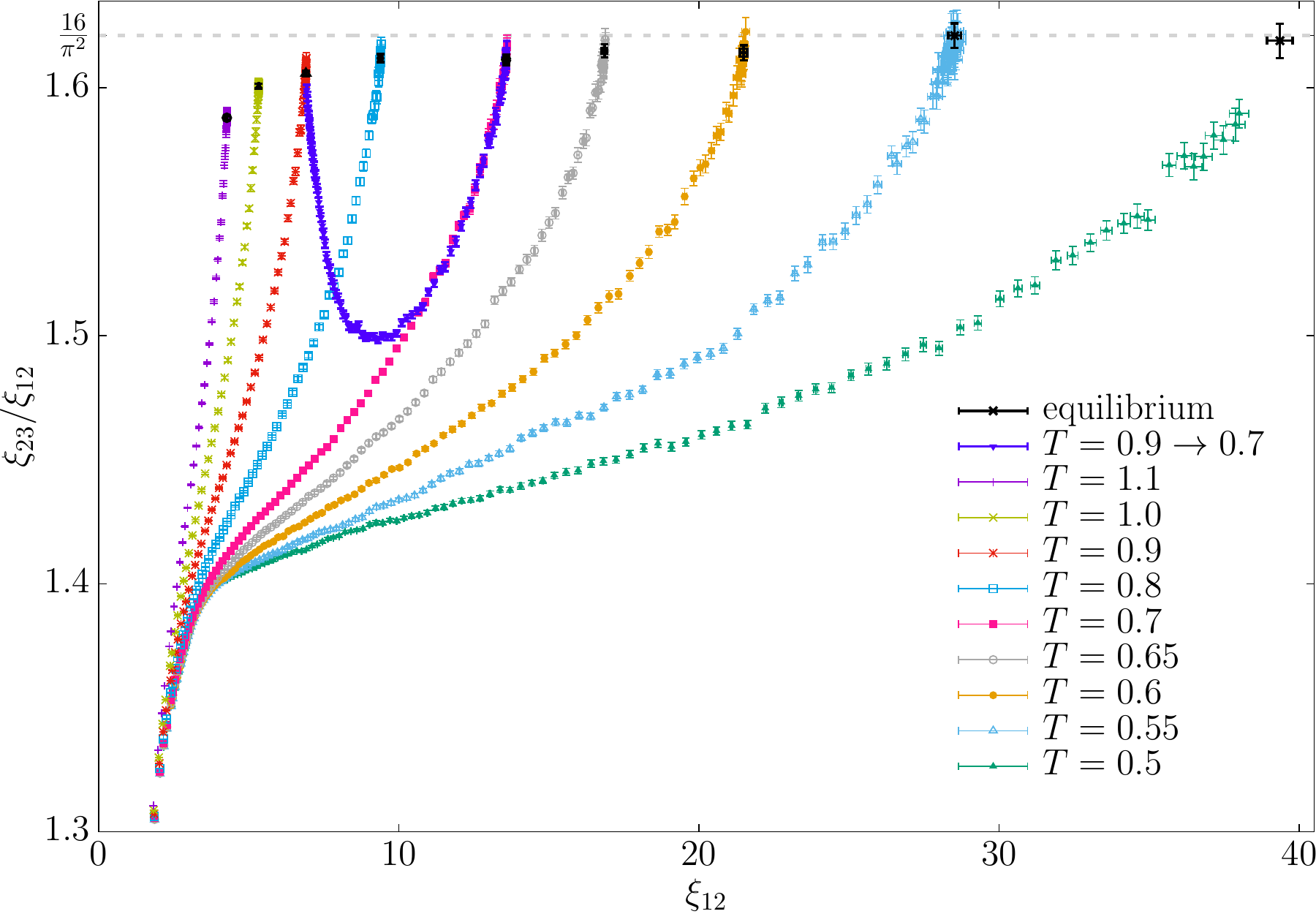}
\end{center}
\caption{As time evolves [i.e. $\xi_{12}(\tw;T)$ grows until it
    reaches its equilibrium value $\xi_{12}^{\mathrm{eq}}(T)$], the
  scale-invariant ratio $\xi_{23}(\tw,T)/\xi_{12}(\tw,T)$ varies. We
  show data for all the temperatures in our simulation. The dashed
  line corresponds to equilibrium in the limit $T\to 0$ (see
  Ref.~\cite{fernandez:18b}).  We also report results from a different
  protocol, in which the system was first equilibrated at $T=0.9$ and
  then placed suddenly at $T=0.7$. This temperature-shift protocol is
  analyzed further in Fig.~\ref{fig:temperature-shift}.}
\label{fig:R12-1}
\end{figure}

\begin{figure}
\begin{center}
\includegraphics[width=\columnwidth,angle=0]{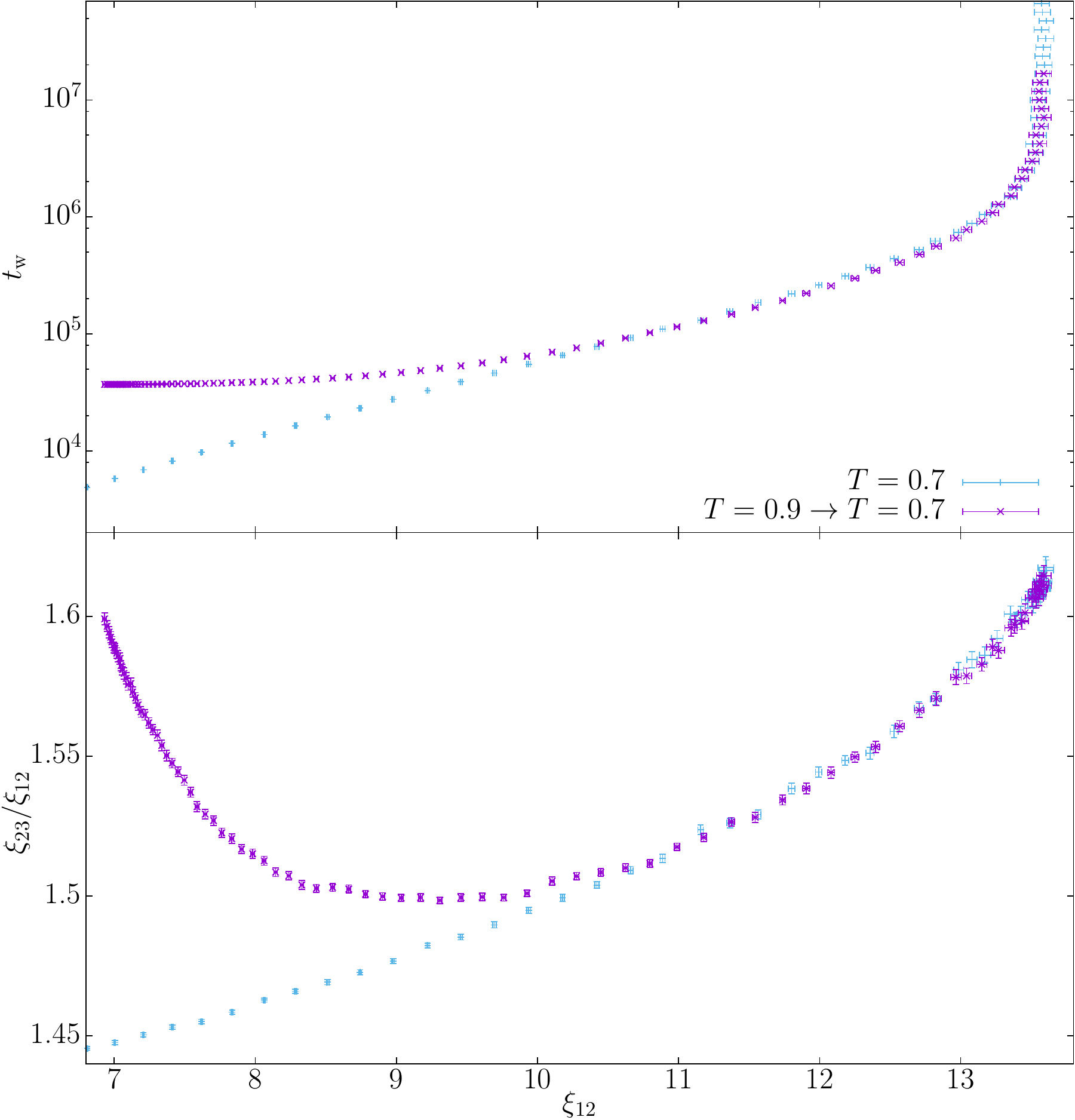}
\end{center}
\caption{{\bf Top:} We compare the growth of the correlation length
  $\xi_{12}$ for the standard run at $T=0.7$ and for the two
  temperature-steps run, that was first equilibrated at $T=0.9$ and then
  placed at $T=0.7$. We set $\tw=0$ for the $T=0.9\rightarrow T=0.7$ run as
  the moment in which temperature reached $T=0.7$. The plot shows $\tw$ as a
  function of $\xi_{12}(\tw;T=0.7)$ (standard run) or $\tw+t_\mathrm{eff}$ as
  a function of $\xi_{12}(\tw;T=0.9\rightarrow T=0.7)$ (two temperature-steps
  run).  Setting the effective time to $t_\mathrm{eff}=3.7\times 10^4$, we
  see that the two curves coincide for $\xi_{12}\gtrsim 10.5$. {\bf Bottom:}
  Zoom of Fig.~\ref{fig:R12-1}, that probes the functional form of the
  correlation function through the scale-invariant ratio
  $\xi_{23}/\xi_{12}$. When the $T=0.7$ and the $T=0.9\rightarrow
  T=0.7$ run match the growth of their respective correlation lengths at
  $\xi_{12}\gtrsim 10.5$, the functional form of their correlation function
  is also matched.}
\label{fig:temperature-shift}
\end{figure}

The out-equilibrium dynamics of the $2D$ Ising spin glass has been
recently studied with an annealing protocol, in which temperature is
slowly varied as time pass by, see Refs.~\cite{xu:17,rubin:17}.
We consider here a numerical experiment in which a
system is first put to thermal equilibrium at $T=0.9$.  When
equilibrium is reached, the system is suddenly placed at $T=0.7$. The
dynamics of the isothermal aging and of the two-steps protocol is
compared in Figs.~\ref{fig:R12-1} and~\ref{fig:temperature-shift}.

We first note, Fig.~\ref{fig:temperature-shift}--top, that the effect of
equilibrating at $T=0.9$ may be aptly described as cumulative
aging~\cite{jonsson:02,bert:04,jimenez:05,maiorano:05}. Indeed, if one neglects a short-time 
transient, the growth of the correlation length for the two temperature-steps
protocol matches the one of the isothermal aging protocol. Equilibrating at
$T=0.9$ translates to the gain of some effective time
$t_\mathrm{eff}=3.7\times 10^4$.

It is also interesting to consider the space-dependence of the
correlation function as probed by the scale invariant ratio
$\xi_{23}/\xi_{12}$, see Figs.~\ref{fig:R12-1}
and~\ref{fig:temperature-shift}--bottom. When placed at $T=0.7$, the system
initially equilibrated at $T=0.9$ effectively \emph{rejuvenates}. As time goes
by, the ratio $\xi_{23}/\xi_{12}$ decreases, rather than
increasing. However, at some point the decreasing $\xi_{23}/\xi_{12}$
catches with the corresponding increasing ratio for the isothermal aging
system. From that point on, the two curves merge and grow again to their
equilibrium value. Interestingly enough, this merging occurs at the same
$\xi_{12}\approx 10.5$ that sets the cumulative-aging regime.

In summary, our simulations suggest that it will be ultimately possible to
analyze the annealing protocol in terms of an effective time, as suggested by
the simple cumulative aging picture~\cite{jonsson:02,bert:04,jimenez:05,maiorano:05}.

%%%%%%%%%%%%%%%%%%%%%%%%%%%%%%%%%%%%%%%%%%%%%%%%%%%%%%%%%%%%%%%%%%%%%%%
\section{Conclusions}\label{sect:conclusions}

We have studied the out-of-equilibrium dynamics of the two dimensional
Edwards-Anderson model with binary couplings. The size of the glassy
domains is characterized by a time-dependent coherence
length $\xi_{12}(\tw)$. We have been able to study the full range of
the dynamics: from the initial transients to the equilibrium through
numerical simulations with a time span of 12 orders of magnitude.

In the limit of low temperatures, where the equilibrium
$\xi_{12}^{\mathrm{eq}}$ becomes large, the growth of $\xi_{12}(\tw)$
is ruled by a single scaling function, that we compute. The argument
of the scaling function is the time elapsed since the quench to the
working temperature, namely $\tw$, as measured in units of a
temperature dependent timescale $\tau(T)$. This result reconciles old
and recent experiments (we also provide rigorous support to the interpretation of those experiments):
\begin{itemize}
\item On the one hand, we show that it is possible to reach
  equilibrium in two-dimensional spin glasses. The notion of a maximal
  barrier $B_\mathrm{max}(T)$ (and, therefore an equilibration time
  $\sim\mathrm{e}^{B_\mathrm{max}(T)/T}$) is central in the analysis
  of experimental spin-glass dynamics in a film
  geometry~\cite{guchhait:14,guchhait:15a,guchhait:15b,guchhait:17,zhai:17}. Our
  scaling function shows that any (sensible) empirical determination
  of the equilibration time will be proportional to the only intrinsic
  timescale in the problem, namely our $\tau(T)$.\footnote{In order to
    avoid confussions, let us recall that dynamics is \emph{not} ruled
    by a single time scale. Quite on the contrary, see
    Eq.~\eqref{eq:Laplace-ideal}, one needs a continuous distribution
    of time scales. Furthermore, this distribution of characteristic
    times is extremely broad [mind the logarithm of $\tau$ in
      Eq.~\eqref{eq:Laplace-ideal}].  What we argue from our data is
    that the temperature evolution of this very complex behaviour can
    be encoded though a temperature dependent $\tau(T)$.} Hence,
  the experimental determination of $B_\mathrm{max}(T)$ is physical.
\item On the other hand, because the only intrinsic time scale in the
  problem $\tau(T)$ grows exceedingly fast upon decreasing
  temperature, recall Eq.~\eqref{eq:barreras-criticas}, it is sensible
  to wonder about the behaviour at times $\tw\ll\tau(T)$. This is
  precisely the case for the old single-layer
  experiments~\cite{schins:93}. In the regime $\tw\ll\tau(T)$, the
  dynamics appears as critical, $\xi_{12}(\tw)\sim \tw^{1/\hat z}$
  with $\hat z\approx 7$ [recall
    Eq.~\eqref{eq:hat-z-def}]. Furthermore, the exponent $\hat z$ is
  large enough to mimic (for a sizeable range of $\tw$) the
  logarithmic behaviour that was assumed in the analysis
  of~\cite{schins:93}.
\end{itemize}
In other words we have characterized completely the dynamics: this new
characterization will be very useful when discussing the physics of
spin glass films in the next generation experiments. In particular,
studying the $D=3$ to $D=2$ crossover in superspin-glass samples with
a film geometry appears as an exciting possibility. Indeed, it has been
possible to study experimentally $\xi_{12}(\tw)$ in 3D superspin
glasses~\cite{nakamae:12}, therefore investigating the effects of a
film geometry appears as an exciting next step.

We have considered as well more complex temperature-change protocols.
Specifically, we have taken the system to thermal equilibrium at a low
temperature, and then suddenly quench it to an even lower temperature.
A careful consideration of the spatial shape of the correlation
function tells us that the effect of the thermalization in the first
temperature-step can be described as cumulative
aging~\cite{jonsson:02,bert:04,jimenez:05,maiorano:05}.

Finally, and given that our data do reach thermal equilibrium, we have
revisited the issue of universality. In particular, we have found a
clear evidence of Universality for the thermal exponent $\nu$ (in
previous work~\cite{fernandez:16b}, this exponent was extremely
difficult to compute for the binary couplings $J=\pm 1$ considered
here).

%%%%%%%%%%%%%%%%%%%%%%%%%%%%%%%%%%%%%%%%%%%%%%%%%%%%%%%%%%%%%%%%%%%%%%%
\section{Acknowledgments}

We thank Raymond Orbach and Jos\'e \'Angel del Toro for discussions.
This project has received funding from the European Research Council
(ERC) under the European Union’s Horizon 2020 research and innovation
program (grant agreement No 694925). We were partially supported by
MINECO (Spain) through Grant Nos. FIS2015-65078-C2, FIS2016-76359-P,
by the Junta de Extremadura (Spain) through Grant No. GRU10158 and
IBI16013 (these four contracts were partially funded by FEDER). Our
simulations were carried out at the BIFI supercomputing center (using
the \emph{Memento} and \emph{Cierzo} clusters) and at ICCAEx
supercomputer center in Badajoz (\emph{Grinfishpc} and
\emph{Iccaexhpc}). We thank the staff at BIFI and ICCAEx
supercomputing centers for their assistance.

\appendix
%%%%%%%%%%%%%%%%%%%%%%%%%%%%%%%%%%%%%%%%%%%%%%%%%%%%%%%%%%%%%%%%%%%%%%%
\section{Details of simulation}\label{sect:simu-details}

We have simulated the Metropolis dynamics of the very same $96$
disorder samples of $L=512$ lattices, at temperatures
$T=0.5, 0.55, 0.6, 0.65,0.7,0.8,0.9,1.0$ and $1.1$.
At every temperature, the simulation has
lasted for as long as needed to reach thermal equilibrium, see
Fig.~\ref{fig:xioverxieq}.  We have stored configurations at times
$\tw=\mathrm{integer-part-of}\, 2^{i/4}$, with $i$ integer. Besides, for
check-pointing purposes, we have also stored configurations every $2^{28}$
Metropolis sweeps. In addition, for $T\geq 0.7$ (where equilibrium is reached
very fast), we found it convenient to store more often configurations evenly
spaced in Monte Carlo time. Our aim was gaining precision in the equilibrium
regime (see below).  The statistical analysis has been performed off-line,
from these stored spin configurations.

An unusual feature of our simulations is that the number of replicas
$N_R$ was large, namely $N_R=256$ for $T>0.5$ and $N_R=264$ for
$T=0.5$ (the reason of the difference is explained below). When computing
the correlation function $C_4(r,\tw)$, Eq.~\eqref{eq:C4-def}, we have
$N_R(N_R-1)/2$ distinct choices for the pair of replica indices. It
has been recently noticed~\cite{janus:18} that, when $L\gg\xi(\tw,T)$,
the choice of a large $N_R$ reduces significantly the statistical
errors in the computation of integrals such
as~\eqref{eq:Ik-def}. However, the simulations at the lowest
temperatures, $T=0.55$ and $0.5$ were somewhat special. At $T=0.55$ we
did simulate $N_R=256$ replicas up to times
$\tw< 2^{32}\approx 4.3\times 10^9$.
However, the equilibrium results were obtained from a
longer simulation, $\tw\leq 2^{34}$, albeit with $N_R=32$ only.  As
for $T=0.5$, the simulation with $N_R=264$ replicas lasted to times
$\tw<2^{35}\approx 3.4\times 10^{10}$. Equilibrium results were
obtained from a $N_R=24$ replicas simulation that reached 
$\tw=3\times 2^{37}\approx 4.1\times 10^{11}$.

We have used an additional trick in order to gain statistics in the
computation of the equilibrium $C_4^\mathrm{eq}(r)$.  Indeed, at some
point our data display no measurable dependence of $\tw$, see
Fig.~\ref{fig:xioverxieq}, implying that thermal equilibrium has been
reached. Yet, it is clear that (in equilibrium only!) there is no
particular reason to take the two replicas in Eq.~\eqref{eq:C4-def} at
the same $\tw$. Therefore, for each pair of replicas, we have also
averaged over pairs of times $({\tw}',{\tw}'')$, with both ${\tw}'$ and
${\tw}''$ larger than the safe equilibration threshold time.

We also performed some sort of time condensation in the (late)
out-equilibrium regime in Fig.~\ref{fig:xioverxieq}, for $T=0.5$ and
$0.55$, in order to gain statistics. At $T=0.55$ the data point shown
at $\tw\approx 2^{32.75}$ is obtained by averaging the overlap for all
pair of times $({\tw}',{\tw}'')$, with ${\tw}'$ and ${\tw}''$ in the
set of 8 times nearest to $2^{32.75}$ (we stored
configurations every $2^{28}$ Monte Carlo time steps). In the case of
$T=0.5$, the data shown in Fig.~\ref{fig:xioverxieq} for
$2^{35}\leq \tw<2^{37}$ where obtained from all pairs $({\tw}',{\tw}'')$, with
${\tw}',{\tw}''\in(2^{-\frac{1}{8}}\tw,2^{\frac{1}{8}}\tw)$. The
horizontal error bars in Fig.~\ref{fig:xioverxieq} span the averaging
interval for each data point. These blurred time data points were used
as well the rest of the figures.

All our simulations employed multispin coding (see~\cite{newman:99}
for a general introduction). Specifically, we adapted to $D\!=\!2$ the
three-dimensional \emph{daemons} algorithm~\cite{ito:90}, which is
extremely sober in the number of Boolean operations requested. Given
that the timescale of the simulation changes by orders of magnitude
upon lowering the temperature, we employed two different simulation
strategies. Multi-replica multispin coding (MURE) was employed for
$T>0.55$. On the other hand, the simulations at $T=0.55$ and $0.5$
were so demanding that we needed to develop a new computer program
implementing the more sophisticated multisite multispin coding
(MUSI)~\cite{fernandez:15}. The MUSI code turned out to be
significantly more effective than the previous MURE program. In fact,
most of the CPU time invested, employed the MUSI code.

In the MURE simulation, each of the 256 bits in a computer word is used to
represent a different real replica. This choice is typically regarded as too
costly, because one needs to generate an independent random number to simulate
every one of the 256 bits. Fortunately, the Gillespie
method~\cite{gillespie:77,bortz:75} (once adapted to multispin coding
simulations~\cite{fernandez:15}) is very efficient in solving this problem at
low temperatures.

In a MUSI simulation (see~\cite{fernandez:15} for a $D=3$ implementation), the
256 bits in a computer word represent 256 spins in the same lattice. We pack
the 256 spins in a single superspin variable. The superspin lattice has a
linear dimension $L/16$. Our chosen correspondence between the physical
coordinates $(x,y)$ and the superspin coordinates  $(i_x,i_y)$ is
\begin{eqnarray}
x&=&b_x\frac{L}{16}+i_x\,,\\
y&=&b_y\frac{L}{16}+i_y\,,\\ 
0&\leq& i_x,i_y<\frac{L}{16}\,,\ 0\leq b_x,b_y< 16\,. 
\end{eqnarray}
In this way, 256 physical coordinates $(x,y)$ are assigned to the very same
superspin coordinates $(i_x,i_y)$. The bit index $0\leq i_b=16 b_y+b_x <256$\,
unambiguously identifies the physical coordinates.

We have employed \emph{Pthreads} to further speed up the MUSI
simulation. Each thread ran a different replica of the same sample. In
this way, all the threads could share the memory for the coupling
matrix $\{J_{\mathitbf{x},\mathitbf{y}}\}$. The $T=0.55$ simulation
was ran in the \emph{Grinfishpc} and \emph{Iccaexhpc} clusters, based
on a AMD Opteron (TM) 6272 processor. In our optimal configuration, 64
threads simulate two independent samples, 32 replicas per sample, at
an overall speed of 2.9 picoseconds/spin-flip. On the other hand, the
$T=0.5$ simulation was ran on Intel(R) Xeon(R) E5-2680v3 processors of
the \emph{Cierzo} cluster. A total of  24
cores are arranged in each \emph{Cierzo}'s dual board (hence the $N_R=24\times 11=264$ replicas simulated for $T=0.5$), and
simulates 24 replicas of the same sample at $1.8$
picoseconds/spin-flip (1.9 picoseconds/spin-flip at
$T=0.55$).\footnote{For sake of completeness, we also give the
  performance figures for MURE multispin coding as measured at
  $T=0.55$. Recall that our MURE code simulates $N_R=256$ replicas
  simultaneously. In the optimal configuration for the Opteron (TM)
  6272 processor, 32 threads collaborate in the simulation of 2
  independent samples (hence 16 threads per sample), at an overall
  speed of 11 picoseconds/spin-flip. On the other hand, 24 threads in
  the Intel(R) Xeon(R) E5-2680v3 simultaneously simulate three
  independent samples at a rate of 10 picoseconds/spin-flip.}

%%%%%%%%%%%%%%%%%%%%%%%%%%%%%%%%%%%%%%%%%%%%%%%%%%%%%%%%%%%%%%%%%%%%%%%
\section{Truncating the Integrals}\label{sect:truncation}

We provide here specific details about our numerical computation of the
integrals $I_k(\tw)$, defined in Eq.~\eqref{eq:Ik-def}, that we repeat
here for the reader's convenience
\begin{equation}\label{eq:app-Ik-def}
  I_k(\tw)=\int_0^{\infty}\mathrm{d}\,r\ r^k C_4(r;\tw)\,.
\end{equation}
The basic observation is that the $C_4(r,\tw)$ function decays very
fast with $r$. At finite $\tw$ the system has no time to generate a
pole-singularity for the Fourier transform of $C_4(r,\tw)$, implying
that $C_4(r,\tw)\sim\mathrm{e}^{-(r/{\hat\xi})^\beta}$ with $\beta>1$
($\beta=1$ in equilibrium~\emph{only}),
see Ref.~\cite{fernandez:18b}. Therefore, the contribution to
the integrals~\eqref{eq:app-Ik-def} of the region $r\gg\xi$ is just
statistical noise. We follow the strategy introduced in
Refs.~\cite{janus:08b,janus:09b} to avoid the noise without incurring
in serious truncation errors. Let us briefly recall the method here,
with some details specific to our implementation.

One start by introducing a noise-induced cutoff, $r_\mathrm{c}$. Let
$\delta C_4(r,\tw)$ be the statistical error in our computation of
$C_4(r,\tw)$.  Then, $r_\mathrm{c}$ is the smallest distance such that
$C_4(r_\mathrm{c},\tw) < 3\,\delta C_4(r_\mathrm{c},\tw)$.

Next, in order to account for the tail of $C_4(r,\tw)$ we perform a fit to (recall that $L=512$)
\begin{eqnarray}\label{eq-app:fit-C4}
  C_4(r,\tw)&=& \frac{a}{\sqrt{r}}\mathrm{e}^{-(r/{\hat\xi})^\beta}\
  +\ [\,r\rightarrow (L-r)\,] \,,\\\nonumber
&&\ r_\mathrm{min}\leq r \leq r_\mathrm{max}\,,
\end{eqnarray}
where the fit parameters are $a$, ${\hat\xi}$ and $\beta$. In the
equilibrium limit, one should rather use
\begin{eqnarray}\label{eq:non-standard-fit}
  C_4^\mathrm{eq}(r,T)&=&{\cal A}(\xi_\mathrm{exp})
  \left[\,K_0\Big(\frac{r}{\xi_{\mathrm{exp}}(T)}\Big)\ +\right.\\
    && \left.K_0\Big(\frac{L-r}{\xi_{\mathrm{exp}}(T)}\Big)\,\right]
  \,\nonumber
\end{eqnarray}
where ${\cal A}(\xi_\mathrm{exp})$ is an amplitude depending on temperature
through $\xi_\mathrm{exp}(T)$. We have included in~\eqref{eq:non-standard-fit}
the first image term, $K_0[(L-r)/\xi_\mathrm{exp}]$ (we use periodic
boundary conditions), as a further control of finite-size effects.

Also in the out-equilibrium fits to Eq.~\eqref{eq-app:fit-C4}, we add
the first image tem to control finite-size effects, as in
Eq.~\eqref{eq:non-standard-fit}. Fortunately, this precaution turns
out to be inconsequential both in the equilibrium and out-equilibrium
cases.  For the equilibrium correlation-function
$C_4^{\mathrm{eq}}(r)$ we have preferred the exact asymptotic form,
given by Bessel's $K_0$ function, see
Eq.~\eqref{eq:non-standard-fit}. Nevertheless,
Eq.~\eqref{eq-app:fit-C4} provides compatible results (albeit with
larger errors).

The distances $r_\mathrm{min}$ and $r_\mathrm{max}$ are fixed in a
self-consistent way. In the first iteration,
$r_\mathrm{max}=r_\mathrm{c}$ and $r_\mathrm{min}=2$. Next, we check
the fit's figure of merit, the diagonal $\chi^2/\mathrm{dof}$. We
increase $r_\mathrm{min}\rightarrow r_\mathrm{min}+1$ until
$\chi^2/\mathrm{dof}<1$. At this point, we check the difference
between $C_4(r_\mathrm{min},\tw)$ and the fitted function. If this
difference is larger than $1.5\, \delta C_4(r_\mathrm{min},\tw)$, then
we set (iteratively) $r_\mathrm{min}\rightarrow r_\mathrm{min}+1$.

Once $r_\mathrm{min}$ is determined, we repeat the fit, setting
$r_\mathrm{max}=L/2$ (this precaution tries to avoid finite-size
artifacts; however the effects on the fit parameters turn out to be
smaller than a tenth of the error bar). 

The next step is a readjustment of the cut-off distance
$r_\mathrm{c}$, as the minimal distance satisfying
$C_4^{\mathrm{fitted}}(r_\mathrm{c},\tw) < 3\,\delta C_4(r_\mathrm{c},\tw)$.

Finally, the integral from 0 to $r_\mathrm{c}$ is computed with the estimated
$C_4(r,\tw)$ and the integral from $r_\mathrm{c}$ to $\infty$ is carried out
with $C_4^{\mathrm{fitted}}(r,\tw)$. Of course, while computing the integral we subtract from $C_4^{\mathrm{fitted}}(r,\tw)$ the contribution of the first-image term.

To compute errors, we use the jackknife
method~\cite{yllanes:11}. Fits and integrals are computed for each
jackknife block (without varying $r_{\mathrm{min}}$,
$r_{\mathrm{max}}$, $r_{\mathrm{c}}$, which are set with our total
statistics).

As for the quadrature method, we interpolate the integrand
$r^kC_4(r,\tw)$ with a cubic spline, which is then integrated exactly.
The cubic-spline is a cubic polynomial for each interval $k<r<k+1$,
$k$ integer, which is integrated exactly by a second-order Gauss-Legendre
method.

\section*{References}

\bibliographystyle{iopart-num}

\end{document}